\documentclass[11pt,a4paper]{article}

\usepackage[utf8]{inputenc}
\usepackage{a4wide}
\usepackage{amsmath}
\usepackage{amsfonts}
\usepackage{amssymb}
\usepackage{mathtools}
\usepackage{xcolor}
\usepackage{enumerate}
\usepackage{xspace}
\usepackage{subcaption}
\usepackage{xparse}
\usepackage{anyfontsize}
\usepackage[thmmarks, amsmath, thref, hyperref]{ntheorem}

\usepackage{graphicx}    
\usepackage{float} 
\usepackage{graphics} 
\usepackage{gensymb}
\usepackage{balance}
\usepackage{amssymb}
\usepackage{color}
\usepackage{mathrsfs}
\usepackage{scalerel}
\usepackage{stackengine,wasysym}   
\usepackage{subcaption}
\usepackage{caption}
\usepackage{scalerel} 

\usepackage{enumitem} 
\newcommand{\subscript}[2]{$#1 _ #2$}
\usepackage{amssymb}
\newtheorem{corollary}{Corollary}[section]
\newtheorem{theorem}{Theorem}[section]
\newtheorem{problem}{Problem}[section]
\newtheorem{remark}{Remark}[section]
\newtheorem{proof}{Proof} 

\def\real{{\mathbb R}}
\def\comp{{\mathbb C}}
\def \j{\mathbf{j}~}
\def \ie{\text{i.e.~}}
\def \etc{\text{etc.}}

\newcommand{\diag}{\mathop{\mathrm{ diag}}}
\newcommand{\bdiag}{\mathop{\mathrm{  bdiag}}}

\definecolor{temp}{rgb}{0.0, 0.5, 0.0}

\newcommand{\khaled}{\textcolor{black}}

\usepackage[pdfauthor={Khaled Laib},%
            pdftitle={Convex LMI opotimization to solve the uncertain 	power 
fow analysis problem},%
            pdftex,bookmarks]{hyperref}

\definecolor{dred}{rgb}{0.5,0,0}
\definecolor{red}{rgb}{0.8,0,0}
\definecolor{blue}{rgb}{0,0,0.5}
\definecolor{green}{rgb}{0,0.3,0}
\definecolor{grey}{rgb}{0.5,0.5,0.5}

\begin{document}

\title{Convex LMI optimization for the uncertain 	power 
flow analysis}

\author{
	Khaled Laib$^1$,
	Anton Korniienko$^2$, Florent Morel$^2$ and Gérard Scorletti$^2$\\[3ex]
	{ \footnotesize 	$^1$ Univ. Grenoble Alpes, CNRS, GIPSA-lab, Control Systems Department,}
	{ \footnotesize  F-38000 Grenoble, France.} \\
     { \footnotesize (\small e-mail: khaled.laib@ gipsa.lab.fr)} \\[1ex]
	{ \footnotesize 	$^2$  Univ. Lyon, École Centrale de Lyon, Laboratoire Ampère,}
	{ \footnotesize  69134 Écully, France.}\\
  { \footnotesize (\small e-mail: $<$anton.korniienko, florent.morel, gerard.scorletti$>$@ec-lyon.fr)} }
\date{}
\maketitle

	\begin{abstract}                
	This paper investigates the uncertain power flow analysis in distribution networks within the context of renewable power resources integration such as wind and solar power. 		The analysis  aims  to bound the worst-case voltage magnitude in any node of the network for a given uncertain power generation  scenario. 		The major difficulty of this problem is the non-linear aspect of power flow equations.  The proposed approach    does not require the linearization of   these equations   and   formulates the problem as an optimization problem with polynomial constraints. A new   tool   to  investigate the feasibility of such  problems  is presented and it  is obtained as an extension of the $\mathcal{S}-$procedure, a fundamental result in robustness analysis. 		A solution to the uncertain power flow  analysis problem is proposed using this new tool. 		The different  obtained  results of this paper are expressed as LMI optimization  problems which guaranties  an efficient numerical resolution  as it will be  demonstrated through an illustrative   example. 
	\end{abstract}



\section{Introduction}  

The integration of renewable power resources such   as wind and solar power into  the  existing  distribution networks\footnote{Power distribution network is the terminal part of power network where residential buildings, schools, \etc~are found.} has become a  necessity  in order to create an  environmental  responsible energy usage. Nevertheless, these renewable power resources  are     intermittent and difficult to  predict  accurately which make them a source of uncertainty in power systems.   This paper focuses on  the effect of this uncertain   power    integration         on the network voltage  magnitudes    by computing  their worst-case  upper and lower bounds for a given renewable power generation scenario. This problem is known as   the uncertain power flow analysis.  

 \medskip

Uncertain power flow analysis  considers the network performance  in steady state by investigating if  the different   voltage magnitude bounds remain  within the acceptable  interval defined by   power system operational requirements. Furthermore, since it   is an off-line analysis, the uncertain power flow analysis is very beneficial in many  operations which do not require fast responses.  For instance,    in  authorizing  further integration  of renewable power resources,   in  scheduling  network interventions and in  defining   power  system operations across different time-scales:  from day-ahead to long period scheduling. 
Therefore, the uncertain power flow analysis has received an important attention over the last decades and it is possible to distinguish two main categories of approaches:  probabilistic and  deterministic.  

\medskip

In probabilistic approaches, e.g.\cite{FVHA:13, BCH:14,ChD:17},  the power generation uncertainty is modeled  as random variable   with predefined distribution functions. Probability theory is used  to  obtain  the  probability  distribution  of  power  flow  solutions. However, using these approaches, no strict voltage magnitude bounds are obtained since the power  flow  solutions are given as probability distributions and hence no  worst-case warranty  can be obtained.

\medskip

In deterministic approaches, the  uncertain  power generation   is characterized using sets such as polytopes and ellipsoids.

\medskip

\khaled{In the case when the generated  (injected) power is  characterized with  polytopes, 	interval methods  can be applied. Theses  methods employ different techniques to deal with the non-linear aspects of power flow equations. For instance,  iterative techniques   in~\cite{LZSWYG:14}  and inclusion analysis     in~\cite{LLZWQ:17}. These approaches have several advantages.  However,   the computation complexity  may be important due to some matrix  interval inversions  at each iteration in~\cite{LZSWYG:14}. Moreover,  the    obtained bounds in~\cite{LLZWQ:17} may be  conservative or even  the set of obtained solutions may be empty  because of those inclusion techniques.  }

\medskip

In the general case when the injected power is  characterized with ellipsoids, see e.g.~\cite{SaS:06}, methods of \cite{CJD:11} can be applied. This approach consists in projecting the injected power ellipsoid into the voltage magnitude set using   a linear model of power flow equations with the assumption that power generation variation is sufficiently small.	
However, because of the performed linearization, the obtained results are local  and only valid around the operating point.

\medskip

This paper focuses on the general case   when  the injected power is  characterized with  ellipsoids. In contrast with~\cite{CJD:11}, the linearization of power flow equations is not required in   our approach and hence large injected power variations are allowed.
We reveal   that solving the uncertain power flow analysis problem requires the resolution of an optimization problem with non-linear constraints. More precisely, the constraints involved in this problem are polynomial.

\medskip

The main contribution of this paper is  Theorem~\ref{thm:feasp_with_polynomial_constraints} which is   a  new  tool to investigate the feasibility of set of polynomial constraints using  convex optimization constrained by  linear matrix inequalities~(LMI), see e.g.~\cite{BEFB:94}. This theorem represents   an extension to the  well-known  $\mathcal{S}-$procedure, see e.g.~\cite{BEFB:94, Ulf:01}, in the case of polynomial constraints with complex variables.  The $\mathcal{S}-$procedure is a fundamental result in robustness analysis: this fact reveals the strong connections between uncertain power flow analysis and usual robustness analysis.  	
Another contribution of this paper is  Corollary~\ref{cor:uncertain_power_flow_cor_1} which is   a new  solution     to   the  uncertain power flow analysis problem.

\subsubsection*{Paper outline} 
This paper is organized as follows. Section~\ref{sec:prob_formulation} presents some  power network preliminaries  followed by formulating  the uncertain power flow analysis problem. 
Section~\ref{sec:proposed_approach} presents  a reformulation of this  problem  within the context of   optimization problems with polynomial constraints.   Section~\ref{sec:main_result} presents the main contribution of this paper  while  its application to solve the uncertain power flow analysis problem  is presented in Section~\ref{sec:main_result_appl}. The efficiency of the proposed solution   is demonstrated  through an illustrative   example in  Section~\ref{sec:Illustration_Example}. Conclusions and perspectives are presented in     Section~\ref{sec:conclusion}.

\bigskip

This paper is the long version of \cite{LKMS:18}. To simplify the presentation, all of the proofs and computation details are given in the appendices.


\subsubsection*{Notations}
$\real$ and $\comp$ are the sets of real and complex numbers respectively. 
$\mathcal{N}$ denotes the finite set  $\{1,\dots, N\}$ and~$\j$ denotes the square root of -1.  The transpose and the transpose conjugate of  $X$ are  denoted  $X^T$ and $X^*$ respectively.  
For several scalars $\tau_i$ (respectively several matrices $Q_i$), $\diag_i(\tau_i)$  (respectively $\bdiag_i (Q_i)$) denotes the  diagonal matrix composed of     $\tau_i$ (respectively $Q_i$).   
$u_k$ is the {$(  N^2+N+1)$}   null row vector     except the $k^\text{th}$  entry which is equal to 1. 
At last and in order to avoid repetitions,  the expression
$  	\left( \star\right) ^* M x$ (respectively $\left( \star\right) ^T M x$) replaces any quadratic form such as $x^* M x$ (respectively $x^T M x$).


\section{Preliminaries and Problem formulation} 
\label{sec:prob_formulation}

\subsection{Preliminaries}
\subsubsection{Generalities on power distribution networks}
Consider  a power distribution   network  with $N$ buses (nodes) connected through electrical  lines.
Each of these    buses  represents a power consumer (residential buildings, schools, \etc).  The slack bus (reference bus)  is denoted bus~0 and is located   upstream of the $N$ bus power distribution network.

We assume the following 
\begin{itemize}
	\item The power  network  three-phases     form a balanced system \ie the three phases have the same magnitude and are phase-shifted in time by one-third of the~period. This assumption is required  in order to boil down the   analysis of the three-phase power network     into the analysis of an equivalent one phase power network.
	\item  The power network steady state is established and  the analysis does not concern the  transient state.  
	\item The bus $k$, with $k\in\{1,\dots, N \}$   is   connected to an uncertain  power resource while the power consumption  at this bus is known. The approach presented in this paper can be easily adapted to    other cases\footnote{Other cases such as  uncertain power consumption or both  powers (generation and consumption) are uncertain. Another case is when  only some   buses are connected to  uncertain generation/consumption power.}.
	\item   The slack   bus voltage    is known  and there are no loads or  renewable power resource devices connected to it. 
\end{itemize}

The quantities to be manipulated in  this paper are
\begin{itemize}
	\item The  network admittance matrix 	{$Y$}  defined as 	
	$$ \begin{matrix}
		&\hspace{-5.5cm}Y=Y^T\in \comp^{\left( N+1\right)\times \left( N+1\right) }\\
		&\hspace{0.2cm}Y_{i,j}=
		\left\lbrace
		\begin{matrix}
		&y_{\ell_i}+\displaystyle \sum_{j=1,j\neq i}^{N+1} y_{ij} &\hspace{-1.5cm}\text{if $i=j$} \\[1ex] 
		&\hspace{-1.5cm}-y_{ij} &\text{ if $i  \neq j$ and $i\sim j$}  \\[1ex] 
		&\hspace{-1.5cm}0 &\hspace{-1cm}\text{otherwise} \\
		\end{matrix}
		\right.  
		\end{matrix}$$
	where $y_{\ell_i}$ and  $y_{ij}$  denote the load  admittance    connected  to bus $i$ and  the   line  admittance between bus~$i$ and  bus~$j$ respectively. The symbol  $i\sim j$ means that   bus~$i$ is connected to bus~$j$.  
	\item $\underline{v}_k$ and $\underline{i}_k$ : the (complex) voltage and  (complex) current at bus $k$ respectively. 
	The network voltages and currents are linked through the admittance matrix \begin{equation} 
		\underline{i}_k =  \sum_{j=1}^{N+1} Y_{(k+1),j}~ \underline{v}_{j-1} 
		\label{eqt:link_voltage_current}
	\end{equation}

	\item $\underline{s}_k={p}_k+\j {q}_k$: the (complex) power $\underline{s}_k$, real power  ${p}_k$ and reactive power ${q}_k$ at bus $k$. The bus complex  power is  linked to its  voltage  and current  through 
	\begin{equation}
		\underline{s}_k = \underline{v}_k~\underline{i}^*_k
		\label{eqt:link_power_voltage_current}
	\end{equation}
	\item $\underline{s}_{g_k} ={p}_{g_k}+\j {q}_{g_k}$: the   (complex) generated  power $\underline{s}_{g_k}$, generated real power ${p}_{g_k}$ and generated reactive power ${q}_{g_k}$ at bus $k$.  
	\item $\underline{s}_{\ell_k} ={p}_{\ell_k} +\j {q}_{\ell_k} $: the  (complex) load     power $\underline{s}_{\ell_k}$, load real power ${p}_{\ell_k}$ and load reactive power ${q}_{\ell_k}$ at bus $k$. 
\end{itemize}

The power at each bus $k$ is balanced between generation (injection)  and load, that is     
\begin{equation}
	\underline{s}_k = \underline{s}_{g_k}-\underline{s}_{\ell_k} 
	\label{eqt:link_power_genereted_consummed}
\end{equation}
Hence, for each bus $k$ and by combining equations~\eqref{eqt:link_voltage_current},~\eqref{eqt:link_power_voltage_current} and  \eqref{eqt:link_power_genereted_consummed}, the power flow equations are given by  
\begin{equation}
	\hspace{-0.05cm}\underline{s}_{g_k}-\underline{s}_{\ell_k} = \underline{v}_k  \left( \sum_{j=1}^{N+1} Y_{(k+1),j} \underline{v}_{j-1} \right)^*, ~ k \in \mathcal{N}
	\label{eqt:power_flow_equations}
\end{equation}
As it can be seen, the power flow equations~\eqref{eqt:power_flow_equations} are non-linear with respect to the different $\underline{v}_k$.

Before presenting the  characterization of injected powers~$\underline{s}_{g_k}$, with $k \in \mathcal{N}$,   an important  phenomenon in electric circuit has to be taken into account. This  phenomenon is the electric current magnitude limitations. 

In an electric circuit and   due to   physical properties of the transmission line, the current magnitude transmitted through this line is limited and cannot  exceed some value. Therefore, the magnitude  of  current $\underline{i}_k$ injected into bus~$k$   cannot  exceed a given value $I_k^{max}$, that is
\begin{equation}
	\label{eqt:current_limitation}
	\left| \underline{i}_k \right| < I_k^{max}, ~~~~~~~ k \in \mathcal{N}
\end{equation}

\subsubsection{Characterization of the injected powers}
As   explained above, the powers generated from renewable power resources are variable and difficult to predict with   precision. According  to the literature, ellipsoids are a general    characterization of these powers, see~\cite{SaS:06}.    \\
An  ellipsoid $\mathcal{S}_g$ is a subset of $\comp^N$ and  is   given by 	\begin{equation}
	\mathcal{S}_g =  { \left\lbrace 
		\begin{pmatrix}
			\underline{s}_{g_1}\\ \vdots\\\underline{s}_{g_N}
		\end{pmatrix}
		\in \comp^N
		\left|
		\begin{matrix}
			&\hspace{-0.0cm}	\left(~~\star~~\right)^* \Psi   \left( 		\underline{S}_{g}- 	\underline{S}^0_{g}\right)<1 \\[1ex]
			& \hspace{-2.5cm}\text{with}\\	&\hspace{-0.0cm}~~\underline{S}_{g}=\begin{pmatrix}\underline{s}_{g_1}&\dots&\underline{s}_{g_N}\end{pmatrix}^T\\
			&\hspace{-0.0cm}~~\underline{S}^0_{g}=\begin{pmatrix}\underline{s}^0_{g_1}&\dots&\underline{s}^0_{g_N}\end{pmatrix}^T
		\end{matrix}
		\right. 
		\right\rbrace } \vspace{-0.25cm}
	\label{eqt:power_gen_car}
\end{equation}
where  	\begin{itemize}
	\item  $\underline{S}^0_{g}=\begin{pmatrix}\underline{s}^0_{g_1}&\dots&\underline{s}^0_{g_N}\end{pmatrix}^T$ is the ellipsoid center  and    $\underline{s}^0_{g_k}$ are the nominal values of  injected powers;
	\item $\Psi\in \comp^{N\times N}$ is a hermitian  matrix describing how far the ellipsoid extends
	in every direction.
\end{itemize}

The main interest of  ellipsoidal  characterization is that   it allows to consider   correlations between   different powers in the network which is not possible with polytopic characterization of~\cite{LLZWQ:17}. 

\subsection{Problem formulation}
In the uncertain power flow analysis, the objective is to determine   bounds on the magnitude of each  $\underline{v}_k$, that is $$
V_k^{min}< \left| \underline{v}_k \right| < V_k^{max}~~~~~~~ k \in \mathcal{N}
$$
such that constraints \eqref{eqt:current_limitation} and  \eqref{eqt:power_gen_car} are respected.

These  $2N$  voltage magnitude bounds   inequalities can be rewritten as 
\begin{equation}
	\left( V_k^{min}\right)^2 <   \underline{v}_k^*~\underline{v}_k  < \left( V_k^{max}\right) ^2~~~~~~~ k \in \mathcal{N}
	\label{eqt:voltage_bounds}
\end{equation}
Constraints~\eqref{eqt:voltage_bounds}  form a hyper-rectangle  in $\real^N$ where  the different  $\left(V_k^{min}\right)^2$ and $\left(V_k^{max}\right)^2$ are its  vertices.    This hyper-rectangle  is denoted $\mathcal{V}$  and is given by $$
~~~~~\mathcal{V}={\left\lbrace  
	\begin{pmatrix}
	\underline{v}_1^*\underline{v}_1\\   \vdots \\ \underline{v}_N^*\underline{v}_N
	\end{pmatrix}
	\in \real^N \left|
	\hspace{-0.05cm} \begin{matrix}
	&\left(V_1^{min}\right) ^2  &\hspace{-0.2cm}<  ~\underline{v}_1^*~\underline{v}_1 &\hspace{-0.1cm}<    \left( V_1^{max}\right)^2\\
	&\vdots  &\hspace{-0.3cm}  ~ \vdots &\hspace{-0.1cm}  ~  \vdots\\
	&\left(V_N^{min}\right) ^2  &\hspace{-0.2cm}<  ~ \underline{v}_N^*~\underline{v}_N &\hspace{-0.1cm}<   \left( V_N^{max}\right)^2
	\end{matrix}
	\right. 
	\right\rbrace}.
$$

Therefore, in order to determine the tightest bounds $V_k^{min}$ $V_k^{max}$, it is required to find   the smallest hyper-rectangle; hence the necessity to define   a size measure.

We adopt in this paper the perimeter $\mathscr{P}$ as a size measure for the hyper-rectangle    $\mathcal{V}$. It is given by $$\mathscr{P}=\vartheta~\left( \sum_{k=1}^{N} \left( V_k^{max}\right)^2-\left(V_k^{min}\right)^2 \right) \vspace{-0.1cm} $$ 
where  $\vartheta$ is a positive scalar which depends on  $N$.

\smallskip

After introducing the different concepts of the uncertain power flow  analysis problem and after clarifying its  objective,   it is  now possible to announce the problem formally.

\smallskip

\begin{problem}
	\label{prb:uncertain_power_flow_form_1}
	Consider a power distribution network with $N$ buses and  $Y$ as its   admittance matrix.\\
	The voltage, current and injected power at bus $k$,  with $k\in\mathcal{N}=\{1,\dots,N\}$, are   $\underline{v}_k$, $\underline{i}_k$ and  $\underline{s}_{g_k}$ respectively.

\smallskip

	Given 
	\begin{itemize}
		\item the voltage   $ \underline{v}_0$ at bus 0  (reference bus);
		\item the  limitation $I_k^{max}$  of  $\underline{i}_k$   at bus  $k$ with  $k\in\mathcal{N}$;
		\item the load power    $\underline{s}_{\ell_k}$    at bus $k$ with  $k\in\mathcal{N}$;
		\item the nominal injected  power $\underline{s}^0_{g_k}$   at bus  $k$  with ${k\in\mathcal{N}}$;
		\item the hermitian matrix  $\Psi \in \comp^{N \times N}$.
	\end{itemize}
	
	\smallskip
	
	Find the different   $\left( V_k^{min}\right)^2$ and  $\left( V_k^{max}\right) ^2$  which  
	$$\min_{\left( V_k^{min}\right)^2,\left( V_k^{max}\right)^2}~~~\mathscr{P}=  \vartheta~\left( \sum_{k=1}^{N} \left( V_k^{max}\right)^2-\left(V_k^{min}\right)^2 \right) $$
	subject to $$\left( V_k^{min}\right)^2 <   \underline{v}_k^*~\underline{v}_k  < \left( V_k^{max}\right) ^2~~~~~~~ k \in \mathcal{N}$$
	for every  $\underline{v}_k$ such that 
	\begin{itemize}
		\item $	{ \hspace{-0.1cm} \begin{pmatrix}
			\underline{s}_{g_1}\\ \vdots\\\underline{s}_{g_N}
			\end{pmatrix}} \hspace{-0.09cm} \in \hspace{-0.07cm} { \left\lbrace 
			\hspace{-0.1cm}\begin{pmatrix}
			\underline{s}_{g_1}\\ \vdots\\\underline{s}_{g_N}
			\end{pmatrix}
			\hspace{-0.05cm}\in \comp^N \hspace{-0.07cm}
			\left| \hspace{-0.05cm}
			\begin{matrix}
			&\hspace{-0.0cm}	\left( ~~\star~~\right)^* \Psi   \left( 		\underline{S}_{g}- 	\underline{S}^0_{g}\right)<1 \\[1ex]
			& \hspace{-2.5cm}\text{with}\\	&\hspace{-0.0cm}~~\underline{S}_{g}=\begin{pmatrix}\underline{s}_{g_1}&\dots&\underline{s}_{g_N}\end{pmatrix}^T\\
			&\hspace{-0.0cm}~~\underline{S}^0_{g}=\begin{pmatrix}\underline{s}^0_{g_1}&\dots&\underline{s}^0_{g_N}\end{pmatrix}^T
			\end{matrix} \hspace{-0.1cm}
			\right. 
			\right\rbrace } \hspace{-0.05cm} ;$\\ 
		
		\item $\left| \underline{i}_k\right| <  I_k^{max}$, for every $k\in\mathcal{N}$.   
	\end{itemize}
	with  	
	\begin{itemize}
		\item	 $\underline{s}_{g_k}=\underline{s}_{\ell_k}+\underline{v}_k \left( \displaystyle \sum_{j=1}^{N+1} Y_{(k+1),j}~ \underline{v}_{j-1} \right)^*;$\vspace{-0.15cm}
		\item $\underline{i}_k =  \displaystyle \sum_{j=1}^{N+1} Y_{(k+1),j}~ \underline{v}_{j-1}.$
		
	\end{itemize}
	
\end{problem}


\section{Proposed approach}  \label{sec:proposed_approach}

The different constraints of  Problem~\ref{prb:uncertain_power_flow_form_1} are given in terms of voltages $\underline{v}_k$, injected powers $\underline{s}_{g_k}$ and currents~$\underline{i}_k$. Therefore, the first step toward  the resolution of Problem~\ref{prb:uncertain_power_flow_form_1} is to  rewrite all of its  constraints   in an explicit form in terms of  voltages $\underline{v}_k$. 


The  injected power constraint     
$$	{   \begin{pmatrix}
	\underline{s}_{g_1}\\ \vdots\\\underline{s}_{g_N}
	\end{pmatrix}} \hspace{-0.09cm} \in \hspace{-0.07cm} { \left\lbrace 
	\hspace{-0.1cm}\begin{pmatrix}
	\underline{s}_{g_1}\\ \vdots\\\underline{s}_{g_N}
	\end{pmatrix}
	\hspace{-0.05cm}\in \comp^N \hspace{-0.07cm}
	\left| \hspace{-0.05cm}
	\begin{matrix}
	&\hspace{-0.0cm}	\left( ~~\star~~\right)^* \Psi   \left( 		\underline{S}_{g}- 	\underline{S}^0_{g}\right)<1 \\[1ex]
	& \hspace{-2.5cm}\text{with}\\	&\hspace{-0.0cm}~~\underline{S}_{g}=\begin{pmatrix}\underline{s}_{g_1}&\dots&\underline{s}_{g_N}\end{pmatrix}^T\\
	&\hspace{-0.0cm}~~\underline{S}^0_{g}=\begin{pmatrix}\underline{s}^0_{g_1}&\dots&\underline{s}^0_{g_N}\end{pmatrix}^T
	\end{matrix} \hspace{-0.1cm}
	\right. 
	\right\rbrace } \hspace{-0.05cm} $$
rewrites  as \\[1ex]
$${\forall~V \in \comp^N~~~~{ \begin{pmatrix}\star\\\star\\~\star~\end{pmatrix}}^*~
	Q^{S_g}~
	{  	\begin{pmatrix}	V\otimes V^{*^T}\\V~~\\1~~	\end{pmatrix} } <0 }$$
where
\begin{itemize}
	\item $V= \begin{pmatrix}\underline{v}_1& \dots    &\underline{v}_N 	\end{pmatrix}^T    \in   \comp^{N};$
	\item ${ V\otimes V^{*^T}= \begin{pmatrix}
		\underline{v}_1\times V^*& \dots &\underline{v}_N\times V^*
		\end{pmatrix}^T   \in    \comp^{N^2} };$ 
	\item $Q^{S_g}$ is a $(N^2+N+1)$ by $(N^2+N+1)$  hermitian matrix and its   expression is given by~\eqref{eqt: Q^{S_g}} in   Appendix~\ref{First_Appendix}.
\end{itemize}



The current magnitude constraints$$\left| \underline{i}_k\right| <  I_k^{max},~~~~~~~k\in\mathcal{N}$$
rewrite  as \\
$${ \forall~ V  \in   \comp^N~~~{ \begin{pmatrix}\star\\\star\\~\star~\end{pmatrix}}^* 
	Q^{I}_k~ 
	{  	\begin{pmatrix}	V\otimes V^{*^T}\\V~~\\1~~	\end{pmatrix} } <0,~~k \in \mathcal{N}}$$\\[1ex]
where $Q^{I}_k$  is a $(N^2+N+1)$ by $(N^2+N+1)$  hermitian matrix and its   expression is given by~\eqref{eqt:cuurent_QI} in  Appendix~\ref{First_Appendix}.

 \bigskip

	In the sequel and in order to ease the notation, the   matrices   $Q^{S_g},Q^{I}_1, \dots, Q^{I}_N$ are collected in the   set $ \mathcal{Q}$ and they will be denoted  $Q_i$ with $i\in\{1,\dots,N+1\}$, that is  
	$$
	\begin{matrix}
	 \mathcal{Q}&=	\left\lbrace Q^{S_g},Q^{I}_1,\dots,Q^{I}_N    \right\rbrace\\
	 &~=	\left\lbrace Q_1,Q_2,\dots,Q_{N+1}    \right\rbrace
	\end{matrix}	
	$$

The    $2N$  constraints   of  \eqref{eqt:voltage_bounds}  rewrite  as 
	$$
\begin{matrix} { \begin{pmatrix}\star\\\star\\~\star~\end{pmatrix}}^*~
Q^{min}_k~
{  	\begin{pmatrix}	V\otimes V^{*^T}\\V~~\\1~~	\end{pmatrix} } >0\\[4ex] { \begin{pmatrix}\star\\\star\\~\star~\end{pmatrix}}^*~
Q^{max}_k~
{  	\begin{pmatrix}	V\otimes V^{*^T}\\V~~\\1~~	\end{pmatrix} } >0 \end{matrix} ~~ k\in\mathcal{N}	
$$	where $Q_k^{min}$ and  $Q_k^{max}$ are      $(N^2+N+1)$ by  $(N^2+N+1)$ symmetric real  matrices  and    their    expressions are given by~\eqref{eqt:Q_min-Q_max}  in Problem~\ref{prb:uncertain_power_flow_form_2} bellow.   

 \bigskip

It is now possible to reformulate Problem~\ref{prb:uncertain_power_flow_form_1} as follows.

\begin{problem}
	\label{prb:uncertain_power_flow_form_2}
	Given the data of  Problem~\ref{prb:uncertain_power_flow_form_1} and    the  set of matrices $\mathcal{Q}=\left\lbrace Q_1,\dots,Q_{N+1} \right\rbrace $. Let the matrices $Q_k^{min}$ and $Q_k^{max}$, with  $k\in\{1,\dots,N\}$, given by \vspace{-0.0cm} \begin{equation}
		\begin{matrix}
			&\hspace{0.2cm} Q_k^{min}={\footnotesize ~~~u_{ \scaleto{N^2+k}{6pt}}^T u_{ \scaleto{N^2+k}{6pt}}-   \left(V_k^{min}\right)^2  u_{ \scaleto{N^2+N+1}{6pt}}^Tu_{ \scaleto{N^2+N+1}{6pt}}}\\[1ex] 
			&\hspace{0.2cm} Q_k^{max}={\footnotesize -u_{ \scaleto{N^2+k}{6pt}}^T u_{ \scaleto{N^2+k}{6pt}}+ \left(V_k^{max}\right)^2  u_{ \scaleto{N^2+N+1}{6pt}}^Tu_{ \scaleto{N^2+N+1}{6pt}}}
		\end{matrix}
		\label{eqt:Q_min-Q_max}
	\end{equation} 		
	Find the different  $\left( V_k^{min}\right)^2 $ and  $\left( V_k^{max}\right)^2 $   which \vspace{-0.1cm}
	$$\min_{\left( V_k^{min}\right)^2,\left( V_k^{max}\right)^2}~~~~\mathscr{P}= \vartheta~\left( \sum_{k=1}^{N} \left( V_k^{max}\right)^2-\left(V_k^{min}\right)^2 \right) $$
	subject to 
	
	 $$\begin{matrix} {~~~~ \begin{pmatrix}\star\\\star\\~\star~\end{pmatrix}}^*~
	Q^{min}_k~
	{  	\begin{pmatrix}	V\otimes V^{*^T}\\V~~\\1~~	\end{pmatrix} } >0~~~\\[4ex]
	~~{ \begin{pmatrix}\star\\\star\\~\star~\end{pmatrix}}^*~
	Q^{max}_k~
	{  	\begin{pmatrix}	V\otimes V^{*^T}\\V~~\\1~~	\end{pmatrix} } >0 \end{matrix} ~~ k\in\{1,\dots,N\}$$
	for every $V$ in $\comp^{N}$ satisfying \\
	$$~~~~~   { \begin{pmatrix}\star\\\star\\~\star~\end{pmatrix}}^*~
	Q_{i}~
	{  	\begin{pmatrix}	V\otimes V^{*^T}\\V~~\\1~~	\end{pmatrix} } <0~~~~~ i\in\{1,\dots,N+1\}$$ \\
	with  	\begin{itemize}
		\item $V= \begin{pmatrix}\underline{v}_1& \dots    &\underline{v}_N 	\end{pmatrix}^T    \in   \comp^{N};$
		\item ${ V\otimes V^{*^T}= \begin{pmatrix}
			\underline{v}_1\times V^*& \dots &\underline{v}_N\times V^*
			\end{pmatrix}^T   \in    \comp^{N^2} }.$
	\end{itemize}
\end{problem}

 \bigskip


In Problem~\ref{prb:uncertain_power_flow_form_2} and due to the non-linear aspect of the power flow equations~\eqref{eqt:power_flow_equations}, developing the  different inequalities    results in a set of polynomial  constraints each of which is of the following form\\
$$ ~~~~~~\displaystyle    \sum_{a=1}^{N} \sum_{b=1}^{N}  \sum_{c=1}^{N}   \sum_{d=1}^{N} \alpha_{ \scaleto{abcd}{5pt}}~\underline{v}^*_a \underline{v}_b  \underline{v}_c  \underline{v}^*_d<0, ~~~~ \alpha_{ \scaleto{abcd}{5pt}} \in \comp.$$

Even without attempting to minimize~$\mathscr{P}$ in Problem~\ref{prb:uncertain_power_flow_form_2}, finding the different  $\left( V_k^{min}\right)^2 $ and  $\left( V_k^{max}\right)^2$  which satisfy  all those polynomial constraints at the same time is a challenging task.  For this reason, we will attempt to solve Problem~\ref{prb:uncertain_power_flow_form_2} in two steps. 
\begin{enumerate}[label=(\alph*)]
	\item Test  if there exist some values (feasible set) of  $\left( V_k^{min}\right)^2 $ and  $\left( V_k^{max}\right)^2 $ for which all the polynomial   constraints (without the cost function) are satisfied.
	\item  Search within the feasible set      for the  values of   $\left( V_k^{min}\right)^2 $ and  $\left( V_k^{max}\right)^2 $ which give the smallest value for the perimeter $\mathscr{P}$.
\end{enumerate}


Testing  the existence of a   set  of values for   $\left( V_k^{min}\right)^2 $ and  $\left( V_k^{max}\right)^2 $ for which all the polynomial   constraints of  Problem~\ref{prb:uncertain_power_flow_form_2} are satisfied is a feasibility   problem. This problem   can be decomposed into $2N$ feasibility   problems each of which consists in testing if 
$$ ~~~~~{ \begin{pmatrix}\star\\\star\\~\star~\end{pmatrix}}^*~
Q_0~
{  	\begin{pmatrix}	V\otimes V^{*^T}\\V~~\\1~~	\end{pmatrix} } >0 ~~~~~~~~~~~~ $$ \\
is respected for every  $V$ in $\comp^N$ satisfying\\
$$ 
~~~~~~~~~~~~~~~~~~~~ { \begin{pmatrix}\star\\\star\\~\star~\end{pmatrix}}^* 
Q_i~
{  	\begin{pmatrix}	V\otimes V^{*^T}\\V~~\\1~~	\end{pmatrix} } <0, ~~  i\in\{1,\dots, N+1\} $$ \\
where $Q_0$ is either equal  to  $Q_k^{min}$ or  $Q_k^{max}$  for a given   $k$ depending on the constraint to be tested. 
\smallskip

We define thus the following  feasibility problem with polynomial constraints.   

\smallskip

\begin{problem} 
	\label{prb:feasp_with_polynomial_constraints}
	Let the $(N^2+N+1)$ by $(N^2+N+1)$ complex hermitian  matrices  $Q_0$ and  $Q_i$, with  $i\in\{1,\dots,N+1\}$.\\
	Test if		
	$$ ~~~~~~{ \begin{pmatrix}\star\\\star\\~\star~\end{pmatrix}}^*
	Q_0~
	{  	\begin{pmatrix}	V\otimes V^{*^T}\\V~~\\1~~	\end{pmatrix} } >0 ~~~~~~~~~~~~ $$  
	is respected for every  $V$ in $\comp^N$satisfying\ 
	$$ 
	~~~~~~~~~~~~~~~~~~~~~ { \begin{pmatrix}\star\\\star\\~\star~\end{pmatrix}}^* 
	Q_i~
	{  	\begin{pmatrix}	V\otimes V^{*^T}\\V~~\\1~~	\end{pmatrix} } <0, ~~  i\in\{1,\dots, N+1\} $$  
	with  	\begin{itemize}
		\item $V= \begin{pmatrix}\underline{v}_1& \dots    &\underline{v}_N 	\end{pmatrix}^T    \in   \comp^{N};$
		\item ${ V\otimes V^{*^T}= \begin{pmatrix}
			\underline{v}_1\times V^*& \dots &\underline{v}_N\times V^*
			\end{pmatrix}^T   \in    \comp^{N^2} }.$
	\end{itemize}
	
\end{problem}

\smallskip

A new tool to solve Problem~\ref{prb:feasp_with_polynomial_constraints} is presented in the next section.  


\section{Main result}  \label{sec:main_result}

Theorem~\ref{thm:feasp_with_polynomial_constraints} which is  the main contribution of this paper is stated in this section.  It gives sufficient conditions to  solve  Problem~\ref{prb:feasp_with_polynomial_constraints} as an optimization problem with LMI constraints.

\smallskip

\begin{theorem}
	\label{thm:feasp_with_polynomial_constraints}
	Given the data  of  Problem~\ref{prb:feasp_with_polynomial_constraints}. Let~$\mathcal{E}$ be the set of hermitian matrices    $\widetilde{Q}_\ell\in\mathcal{E}$ given by \vspace{-0.0cm}
	\begin{equation}
		\mathcal{E}= \left\lbrace  \widetilde{Q}_\ell~~\left|~~  \widetilde{Q}_\ell ~ \in ~\mathcal{E}^1  \cup \mathcal{E}^2  \cup \mathcal{E}^3  \cup \mathcal{E}^4  \cup \mathcal{E}^5  \right. \right\rbrace \vspace{-0.1cm}
		\label{eqt:equ_constraints_matrix_set}
	\end{equation}
	where  
	\begin{itemize}
		\item $\hspace{-0.0cm}\mathcal{E}^1\hspace{-0.07cm}=\hspace{-0.07cm}{\small \left\lbrace \hspace{-0.07cm}
			\widetilde{Q}_\ell\left| \hspace{-0.cm}
			\begin{matrix}
			&\hspace{-1.5cm}\exists~(a,b,c,d) \in \mathcal{N} \times \mathcal{N} \times  \mathcal{N} \times \mathcal{N}  \\ 
			&\widetilde{Q}_\ell=
			\begin{pmatrix}	\star \\ \star\\ \star \\ ~\star~ \end{pmatrix}^T	\hspace{-0.1cm}
			\begin{pmatrix}
			~0&~1&~0&~~0\\~1&~0&~0&~~0\\~0&~0&~0&-1\\~0&~0&\hspace{-0.1cm}-1&~~0
			\end{pmatrix}
			\begin{pmatrix}	u_{(a-1)N}\\u_{(c-1)N+d}\\u_{(d-1)N+b}\\u_{(c-1)N+a}\end{pmatrix}
			\end{matrix}
			\hspace{-0.2cm}
			\right. 
			\right\rbrace }
		$;\\[3ex]

		\item $\hspace{-0.0cm}\mathcal{E}^2\hspace{-0.07cm}=\hspace{-0.07cm}{\small \left\lbrace \hspace{-0.07cm}
			\widetilde{Q}_\ell\left| \hspace{-0cm}
			\begin{matrix}
			&\hspace{-2.5cm}\exists~(a,b,c) \in\mathcal{N} \times \mathcal{N} \times  \mathcal{N}  \\ 
			&\widetilde{Q}_\ell=
			\begin{pmatrix}	\star  \\\star\\ \star \\ ~\star~ \end{pmatrix}^T
			\hspace{-0.1cm}
			\begin{pmatrix}
			~0&~1&~0&~~0\\~1&~0&~0&~~0\\~0&~0&~0&-1\\~0&~0&\hspace{-0.1cm}-1&~~0
			\end{pmatrix}
			\begin{pmatrix}	u_{N^2+a}\\u_{(b-1)N+c}\\u_{N^2+c}\\u_{(b-1)N+c}\end{pmatrix}
			\end{matrix}
			\hspace{-0.15cm}
			\right. 
			\right\rbrace }$;\\[3ex]
		
		\item $\hspace{-0.0cm}\mathcal{E}^3\hspace{-0.07cm}=\hspace{-0.07cm}{\small \left\lbrace \hspace{-0.07cm}
			\widetilde{Q}_\ell\left| \hspace{-0cm}
			\begin{matrix}
			&\hspace{-3cm}\exists~(a,b) \in   \mathcal{N} \times \mathcal{N} \\ &\widetilde{Q}_\ell=
			\begin{pmatrix}	\star \\ \star \\ ~\star~ \end{pmatrix}^T
			\hspace{-0.1cm}
			\begin{pmatrix}
			~~0&~~1&-1\\~~1&~~0&~~0\\-1&~~0&~~0
			\end{pmatrix}
			\begin{pmatrix}	u_{N^2+N+1}\\u_{(a-1)N+b}\\u_{(b-1)N+a} \end{pmatrix}\end{matrix}
			\hspace{-0.15cm}
			\right. 
			\right\rbrace}$;\\[3ex]
		
		\item 
		$ \hspace{-0.0cm} \mathcal{E}^4\hspace{-0.07cm}=\hspace{-0.07cm}{\small \left\lbrace \hspace{-0.07cm}
			\widetilde{Q}_\ell\left| \hspace{-0cm}
			\begin{matrix}
			& \hspace{-4cm}	\exists~a \in \mathcal{N} \\ 	&\widetilde{Q}_\ell=
			\begin{pmatrix}	\star \\ ~\star~ \\ \star\end{pmatrix}^T
			\hspace{-0.1cm}
			\begin{pmatrix}
			~~0&-1&~~0\\~~0&~~0&~~0\\-1&~~0&~~2
			\end{pmatrix}
			\begin{pmatrix}	u_{N^2+N+1}\\u_{(a-1)N+a}\\u_{N^2+a} \end{pmatrix}
			\end{matrix} \hspace{-0.15cm}
			\right. 
			\right\rbrace }
		$;\\[3ex]
		
		\item $ \hspace{-0.cm}
		\mathcal{E}^5\hspace{-0.07cm}=\hspace{-0.07cm}{\small \left\lbrace \hspace{-0.07cm}
			\widetilde{Q}_\ell\left| \hspace{-0cm}
			\begin{matrix}
			&\hspace{-3.2cm}	\exists~a \in \mathcal{N}\\ &\widetilde{Q}_\ell=
			\begin{pmatrix} \star \\ ~\star~ \end{pmatrix}^T
			\hspace{-0.1cm}
			\begin{pmatrix} ~~0 & ~~\mathbf{j}\\ -\mathbf{j} &~~0 \end{pmatrix}
			\begin{pmatrix} u_{(a-1)N+a} \\ 	u_{N^2+N+1} \end{pmatrix}
			\end{matrix}
			\hspace{-0.15cm}
			\right. 
			\right\rbrace}
		$.
	\end{itemize}
	
	The constraint 
	$${ \begin{pmatrix}\star\\\star\\~\star~\end{pmatrix}}^*
	Q_0~
	{  	\begin{pmatrix}	V\otimes V^{*^T}\\V~~\\1~~	\end{pmatrix} }>0 ~~~~~~~~~~~~ $$ \\
	is respected for every  $V$ in $\comp^N$ satisfying
	$$ 
	~~~~~~~~~~~~~~ { \begin{pmatrix}\star\\\star\\~\star~\end{pmatrix}}^* 
	Q_i~
	{  	\begin{pmatrix}	V\otimes V^{*^T}\\V~~\\1~~	\end{pmatrix} } <0, ~~  i\in\{1,\dots, N+1\} $$ \\
	if there exist  $N+1$ positive scalars $\tau_i$  and $N^E$ scalars $\widetilde{\tau}_\ell$ such that \vspace{-0.2cm}
	\begin{equation}
		Q_0 + \displaystyle \sum_{i=1}^{N+1} \tau_i~ Q_i + \displaystyle \sum_{\ell=1}^{N^E} \widetilde{\tau}_\ell~ \widetilde{Q}_\ell >0.
		\label{eqt:main_eqt_thm_feasp_with_polynomial_constraints}
	\end{equation}
	with
	$N^E=N^4+N^3+N^2+2N$ and $\widetilde{Q}_\ell\in\mathcal{E}.$ 
\end{theorem} 


Finding the positive  scalars  $\tau_1, \dots, \tau_{N+1}$ and the scalars  $\widetilde{\tau}_1, \dots, \widetilde{\tau}_{N^E}$ which satisfy  constraint~\eqref{eqt:main_eqt_thm_feasp_with_polynomial_constraints} is  a feasibility problem subject to  LMI constraints. This problem is convex and can be solved efficiently, see~\cite{BEFB:94}.

\smallskip  

\begin{proof}
 See Appendix~\ref{Third_Appendix} 
\end{proof}

\smallskip 

\begin{remark}
	Theorem~\ref{thm:feasp_with_polynomial_constraints} represents an extension    to the well-known $\mathcal S$-procedure, see    \cite{BEFB:94, Ulf:01},  in the case of  polynomial constraints and   with complex variables. For  a set of   Quadratic Constraints~(QC),  and Integral Quadratic Constraints~(IQC) in general, the $\mathcal S$-procedure is used to test if     $X^T  Q_0 X > 0 $ is respected
	for every  $X \in \real^N$ satisfying 
	$   X^T	Q_i X < 0$ with $  i\in\{1,\dots,N+1\}$. The $\mathcal S$-procedure allows to perform this test by finding $\tau_i$ positive scalars with $  i\in\{1,\dots,N+1\}$ such that $	Q_0 +  \sum_{i=1}^{N+1} \tau_i Q_i>0$.   Nevertheless, this result is only valid  when all  components of $X$ are independent which is not the case with the  vector $X={  \left( \begin{smallmatrix} \left( V\otimes V^{*^T }\right)^T~V^T~1 	\end{smallmatrix}\right) ^T}$.  Theorem~\ref{prb:feasp_with_polynomial_constraints} represents an extension  for the $\mathcal S$-procedure  by introducing the scalars $ \widetilde{\tau}_\ell$ and the matrices $ \widetilde{Q}_\ell$ such that  	$Q_0 +  \sum_{i=1}^{N+1} \tau_i Q_i +  \sum_{\ell=1}^{N^E} \widetilde{\tau}_\ell \widetilde{Q}_\ell >0$ where the matrices $ \widetilde{Q}_\ell$  characterize  important  links between the components of $X$, see    Appendix~\ref{Second_Appendix} for the details.  
\end{remark} 

\smallskip

\begin{remark}
	Theorem~\ref{thm:feasp_with_polynomial_constraints} represents an alternative to Sum Of Square (SOS) techniques, see~\cite{Par:03a}, which can be used  to obtain  the different  links between the components of $X={  \left( \begin{smallmatrix} \left( V\otimes V^{*^T }\right)^T~V^T~1 	\end{smallmatrix}\right) ^T}$. In this case, the number    $N^E$  of the  these links  is     $  \frac{1}{2} \frac{\left(k+m \right)! }{k!m!}  \left( \frac{\left(k+m \right)! }{k!m!} + 1 \right)-\frac{1}{2} \frac{\left(2m+k \right)! }{k!m!} $ where $m=4$ and  $k=2(N^2+N)$, see \cite{Par:03a}. Therefore, SOS techniques will be time consuming  when solving Problem~\ref{prb:feasp_with_polynomial_constraints} due to the important number of decision variables.
	Theorem~\ref{thm:feasp_with_polynomial_constraints} represents an alternative  by considering only important links between the components of $X$ and the resulting 
	$N^E$ is equal to $N^4+N^3+N^2+2N$. The result will be an important  reduction  in computation time since the number  of decision variables is significantly reduced.  
\end{remark}


\section{Application to the Uncertain Power Flow Analysis Problem}  \label{sec:main_result_appl}

As stated above in Section~\ref{sec:proposed_approach}, the major difficulty in   Problem~\ref{prb:uncertain_power_flow_form_2}  is its  polynomial     constraints due to the non-linear aspect of the power flow equations~\eqref{eqt:power_flow_equations}.  After proposing  Theorem~\ref{thm:feasp_with_polynomial_constraints} as a new tool to test the feasibility of a set of polynomial constraints, we present  in this section Corollary~\ref{cor:uncertain_power_flow_cor_1} as a new solution to the uncertain power flow analysis problem.

\smallskip

Let $\mathscr{P}_{opt}$ be the optimal perimeter of  Problem~\ref{prb:uncertain_power_flow_form_2}. An upper bound $\widetilde{\mathscr{P}_{opt}}$ on~$\mathscr{P}_{opt}$ can be found using the following corollary.  

\begin{corollary}
	\label{cor:uncertain_power_flow_cor_1}
	Given the data of   Problem~\ref{prb:uncertain_power_flow_form_2}  and let    $\mathcal{E}$  be the set of matrices   $\widetilde{Q}_\ell$ given by~\eqref{eqt:equ_constraints_matrix_set}.   
	Let $N^E=N^4+N^3+N^2+2N$.
	
	\smallskip
	
	An upper bound  $\widetilde{\mathscr{P}_{opt}}$ on the optimal bound of   Problem~\ref{prb:uncertain_power_flow_form_2} can be obtained by finding  for every $k\in\{1,\dots, N\}$ the scalars
	\begin{itemize}
		\item  $\left( V_k^{min}\right)^2 $ and  $\left( V_k^{max}\right)^2$;
		\item $\left( \tau_{min}\right)^k_i $ and  $\left( \tau_{max}\right)^k_i$  with $i\in\{1,\dots, N+1 \}$;
		\item $\left( \widetilde{\tau}_{min}\right)^k_\ell $ and  $\left(\widetilde{\tau}_{max}\right)^k_\ell $ with $\ell\in\{1,\dots, N^E \}$.
	\end{itemize}
	which minimize
	$$~~~~~~~~~~\textnormal{trace } \displaystyle \left( \diag_{k=1,\dots,N}\left( {\left( V_k^{max}\right)^2}-{\left( V_k^{min}\right)^2} \right)\right) $$
	subject to
	\begin{enumerate}[label=(\roman*)]
		\addtolength{\itemindent}{-0.05cm}
		\item ${ \hspace{-0.2cm} \displaystyle \bdiag_{k=1,\dots,N}\left(\hspace{-0.1cm}  Q_k^{min} \hspace{-0.1cm}+ \hspace{-0.1cm}\sum_{i=1}^{N+1} \left( \tau^{min}\right)^k_iQ_i\hspace{-0.1cm} +  \hspace{-0.1cm}\sum_{\ell=1}^{N^E} \left( \widetilde{\tau}^{min}\right)^k_\ell  \widetilde{Q}_\ell\hspace{-0.1cm} \right)\hspace{-0.1cm} >0; }$\\[1ex] \vspace{-0.1cm}
		\item ${ \hspace{-0.2cm} \displaystyle \bdiag_{k=1,\dots,N}\left(\hspace{-0.1cm}  Q_k^{max} \hspace{-0.1cm}+ \hspace{-0.1cm}\sum_{i=1}^{N+1} \left( \tau^{max}\right)^k_iQ_i\hspace{-0.1cm} +  \hspace{-0.1cm}\sum_{\ell=1}^{N^E} \left( \widetilde{\tau}^{max}\right)^k_\ell  \widetilde{Q}_\ell\hspace{-0.1cm} \right)\hspace{-0.1cm} >0; }$\\[1ex] \vspace{-0.1cm}
		\item ${ \displaystyle \hspace{-0.2cm} \diag_{k=1,\dots,N}\left(  \diag_{i=1,\dots,N+1}\left(
			\left( \tau^{min}\right)^k_i
			\right) \right)>0;} $ 
		\item ${ \hspace{-0.2cm} \displaystyle \diag_{k=1,\dots,N}\left(  \diag_{i=1,\dots,N+1}\left(		\left( \tau^{max}\right)^k_i		\right) \right)>0;} $ 

		\item    ${ \hspace{-0.2cm} \displaystyle \diag_{k=1,\dots,N}\left( \diag\left( \left( V_k^{min}\right)^2 , \left( V_k^{max}\right)^2 \right)\right) >0;} $ 
		\item    ${ \hspace{-0.2cm}\displaystyle \diag_{k=1,\dots,N}\left( \left( V_k^{max}\right)^2 -\left( V_k^{min}\right)^2  \right)>0. }$
	\end{enumerate}
	The upper bound  $\widetilde{\mathscr{P}_{opt}}$ is given  by  \\
	${ ~\widetilde{\mathscr{P}_{opt}}=\vartheta~\textnormal{argmin  } \textnormal{trace } \left( \displaystyle \diag_{k=1,\dots,N}\left( {\left( V_k^{max}\right)^2}-{\left( V_k^{min}\right)^2} \right)\right)}  $\\
	such that conditions (i), (ii), (iii), (iv), (v)  and (vi) are respected.
\end{corollary}
\smallskip 
\begin{proof}
	See Appendix~\ref{Fourth_Appendix} 
\end{proof}
\smallskip 
Minimizing   the trace of ${\footnotesize \displaystyle \diag_{k=1,\dots,N}\left( {\left( V_k^{max}\right)^2}-{\left( V_k^{min}\right)^2} \right)}$ in Corollary~\ref{cor:uncertain_power_flow_cor_1}  subject to conditions $(i)$, $(ii)$, $(iii)$, $(iv)$, $(v)$  and $(vi)$ is a problem of minimizing a linear cost function  subject to  LMI constraints. This problem is convex and can be solved efficiently, see~\cite{BEFB:94}. 

\smallskip

\smallskip

In the next section, we  demonstrate the efficiency of our proposed solution through an illustrative example.


\section{Illustration Example}  \label{sec:Illustration_Example}
\begin{figure}[h]
	\centering
	\includegraphics[width=.7\linewidth]{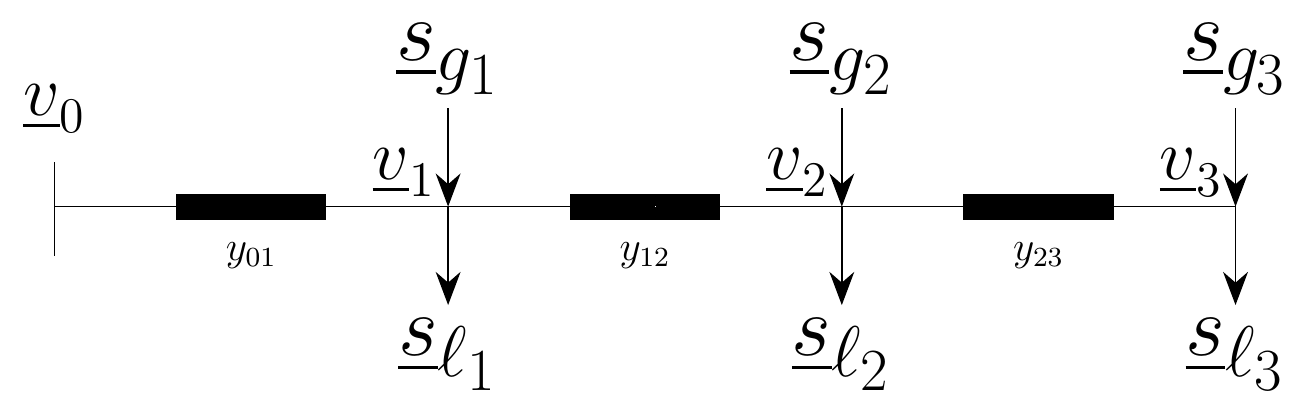}  
	\caption{Example of a 3 bus distribution network.}
	\label{fig:three_nodes_exp}
\end{figure}

We consider a  3 bus distribution network  with injected and load powers~$\underline{s}_{g_k}$ and~$\underline{s}_{\ell_k}$   at each bus $k$
as shown in Fig.\ref{fig:three_nodes_exp}.   This example and its numerical data are taken from~\cite{CJD:11}.  

In this example, none of the  renewable power resources     inject reactive  power into the network, that is $q_{g_k}=0$ with $k\in\{1,2,3\}$.

The data  are normalized and given per unit
\begin{itemize}
	\item the voltage   $\underline{v}_0$ is equal to $0.995~e^{\mathbf{j}0\degree}$;
	\item the load powers  $   \underline{s}_{\ell_1}$, $\underline{s}_{\ell_2}$ and $\underline{s}_{\ell_3}$ are  $0.8+0.25\mathbf{j}$, $0.5+0.1\mathbf{j}$ and $0.9+0.5\mathbf{j}$ respectively;
	\item the current magnitude limitations $I_1^{max}$, $I_2^{max}$ and $I_3^{max}$ are 0.48,   0.23  and   0.66  respectively;
	\item the nominal values of the three voltages   are denoted $\underline{v}^0_1$, $\underline{v}^0_2$ and $\underline{v}^0_3$;	   and     are equal to  $0.987~e^{-\mathbf{j}0.124\degree}$, $0.972~e^{-\mathbf{j}0.273\degree} $ and $0.965~e^{-\mathbf{j}0.302\degree}$ respectively.
\end{itemize}

The injected power vector $\underline{S}_{g}=\left( \underline{s}_{g_1}~~\underline{s}_{g_2}~~\underline{s}_{g_3}\right)^T$   belongs to the  ellipsoid $\mathcal{S}_g $ given by\\
$~~~~~~~~~~~~ \mathcal{S}_g =  { \left\lbrace 
	\underline{S}_{g} 
	\in \comp^3
	\left|~
	\left(~~\star~~\right)^* \Psi   \left( 		\underline{S}_{g}- 	\underline{S}^0_{g}\right)<1 
	\right. 
	\right\rbrace }
$\\
where  $\Psi={  \left( \diag\left(0.08^2,~ 0.06^2,~0.1^2 \right) \right)^{-1}}$ and $\underline{S}^0_{g}=\left( 0.4~~0.3~~0.5\right)^T.$    

\begin{figure}[h]
	\centering
	\begin{subfigure}[b]{0.8\textwidth}	\centering
		\includegraphics[width=0.9\linewidth]{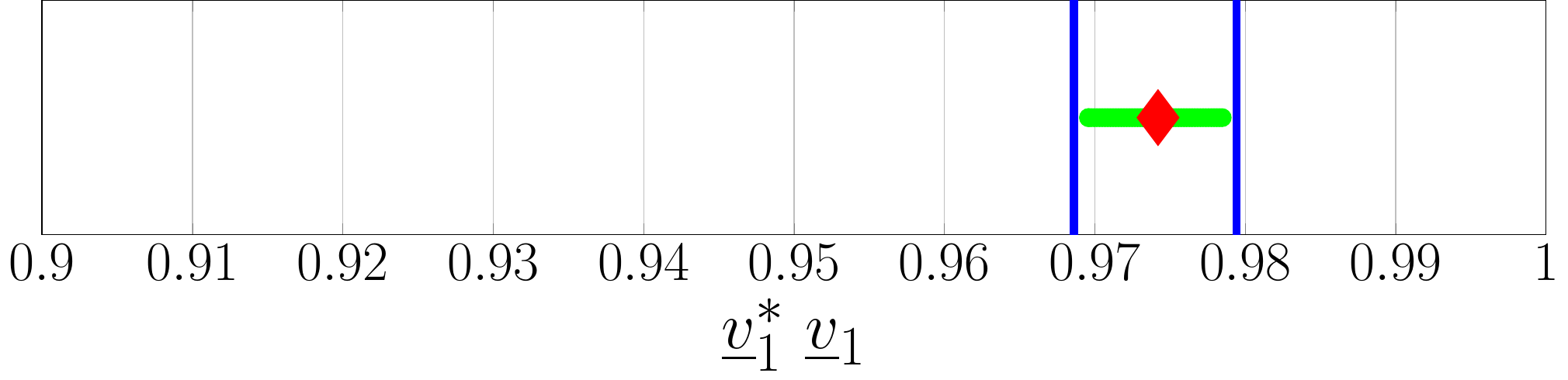}  
	\end{subfigure}\\[1ex]
	\begin{subfigure}[b]{0.8\textwidth}	\centering
		\includegraphics[width=0.9\linewidth]{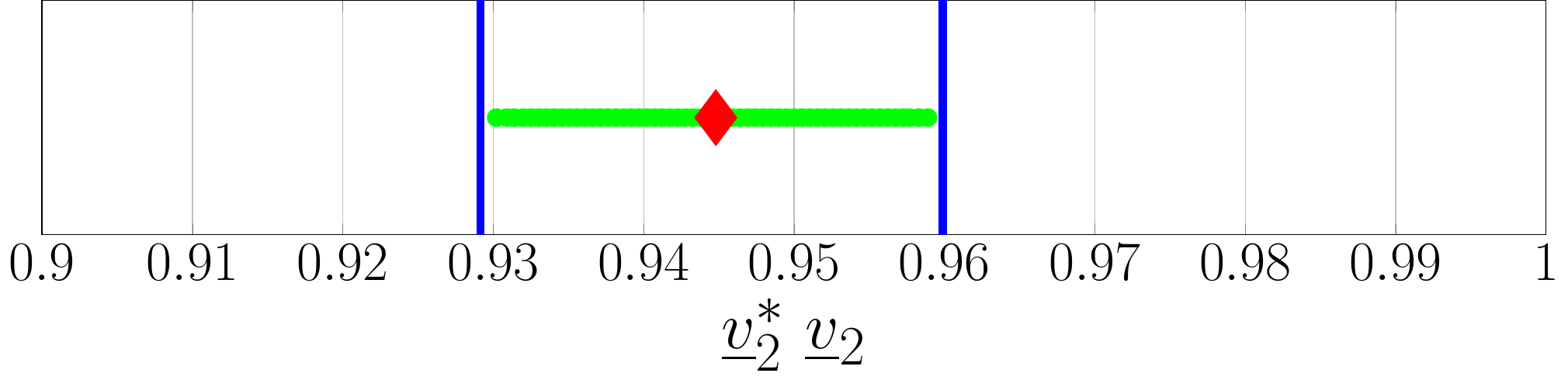} 	 
	\end{subfigure}\\[1ex]
	\begin{subfigure}[b]{0.8\textwidth}	\centering
		\includegraphics[width=0.9\linewidth]{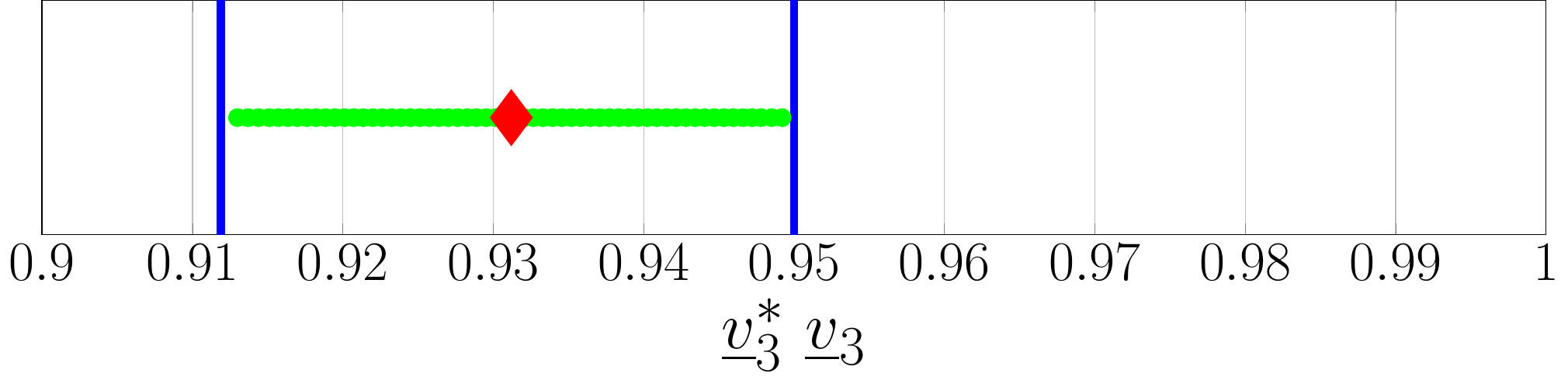}  	 
	\end{subfigure}	  
	\caption{Visualization   of the sampling of  $\protect\underline{v}_k^*~ \protect\underline{v}_k$, (green),     $\left( V_k^{min}\right)^2$ and $\left( V_k^{max}\right)^2$  (blue);  and     $\left( \protect\underline{v}_k^0~\right)^*\protect\underline{v}_k^0$   (red). \label{fig:three_nodes_exp_results} }
\end{figure}

Corollary~\ref{cor:uncertain_power_flow_cor_1} is applied  to find the square of the different lower and upper bounds $V_k^{min}$ and $V_k^{max}$ with~${k\in\{1,2,3\}}$. The results are presented in Fig.~\ref{fig:three_nodes_exp_results} where  
\begin{itemize}
	\item the green dots represent  a sampling of the   variation intervals of  $\protect\underline{v}_k^*~ \protect\underline{v}_k$ such that the different constraints on the injected powers and currents are respected;
	\item the blue  lines represent the different $\left( V_k^{min}\right)^2$ and $\left( V_k^{max}\right)^2$;
	\item the red diamond shapes represent the different $\left( \protect\underline{v}_k^0~\right)^*\protect\underline{v}_k^0$. 
\end{itemize}

The obtained results present few conservatism as  shown in Fig.~\ref{fig:three_nodes_exp_results} and it is possible to obtain the following bounds
$$\begin{matrix}
0.9842 & <& \left| \underline{v}_1\right| &<& 0.9896     \\
0.9639& <&\left| \underline{v}_2\right| &<&0.9797      \\
0.9549& <&\left| \underline{v}_3\right| &<& 0.9747    \\ 
\end{matrix}$$ 
which demonstrates the efficiency of the proposed solution.

\medskip

For comparison,    the obtained results in~\cite{CJD:11} were given as an ellipsoid containing all the voltage magnitudes and independent bounds  cannot be obtained directly while in our approach it is possible to obtain independent bounds directly. Furthermore, the obtained  results  of~\cite{CJD:11} are only valid around the operating point  while our  results do not depend on the operating point since no linearization is required in our approach.


\section{Conclusion}\label{sec:conclusion}
In this paper, the uncertain power flow  analysis problem is investigated. The major difficulty in this problem is the non-linear aspects of the power flow equations. To overcome this difficulty, and to avoid solving the problem locally around an operating point,  our approach reformulates  the problem as an optimization problem with polynomial constraints.
The main contribution of this paper was proposing a new  tool to  solve  the feasibility problem of set of polynomial constraints. Another contribution  was proposing a new   solution to the uncertain power flow analysis problem.
The efficiency of this solution is illustrated through an illustrative example.

As perspective  to this work,  we propose the application of our result on large power network data, see e.g.~\cite{MaP:15}, in order to validate the efficiency of our results on large scale networks.


 \appendix
\section{Appendices}

\subsection{Rewriting the injected power and current magnitudes     constraints of Problem~\ref{prb:uncertain_power_flow_form_1}}
\label{First_Appendix}
The objective of this appendix  is  to rewrite the injected power and current magnitudes     constraints of Problem~\ref{prb:uncertain_power_flow_form_1}  in an explicit form with respect to the   voltages $\underline{v}_k$.

\subsubsection{Rewriting the injected power constraint}  
Using power flow   equations \eqref{eqt:power_flow_equations}, the term $\underline{s}_{g_k}-\underline{s}^0_{g_k}$  is given by
$$\underline{s}_{g_k}-\underline{s}^0_{g_k}={  \displaystyle \sum_{j=2}^{N+1} Y_{(k+1),j}^*\underline{v}_{j-1}^* \underline{v}_k +  Y^*_{(k+1),1} \underline{v}^0_0~\underline{v}_k +\underline{s}_{l_k}-\underline{s}^0_{g_k}} $$
which can be rewritten  as
$$\underline{s}_{g_k}-\underline{s}^0_{g_k}={ \tiny   \left( 
	Y^*_{k+1,2:N+1} ~\vline ~ Y^*_{(k+1),1}\underline{v}^*_0 ~\vline ~ \underline{s}_{l_k}-\underline{s}^0_{g_k}
	\right)} \begin{pmatrix} \underline{v}_k \underline{v}^*_1 \\ \vdots \\ \underline{v}_k \underline{v}^*_N\\ \hline \underline{v}_k \\\hline 1  \end{pmatrix}  $$
where $Y^*_{k+1,2:N+1}$ is the $\left( k+1\right)^\text{th} $ row of the admittance matrix $Y$ taken between  columns 2 and $N+1$. \\[3ex]
The vector $\underline{S}_{g}-\underline{S}_{g}^0$ in the injected power constraint~\eqref{eqt:power_gen_car}  rewrites then  as
$$ ~~~~\underline{S}_{g}-\underline{S}_{g}^0=M_{S_g} {  	\begin{pmatrix}	V\otimes V^{*^T}\\V~~\\1~~	\end{pmatrix} }
$$\\
where \vspace{-0.25cm}
$$M_{S_g}= {  \left(  \bdiag_{\scaleto{k=1,\dots,N}{4pt}}\left(   Y^*_{k+1,2:N+1} \right) ~~ \diag_{\scaleto{k=1,\dots,N}{5pt}} \left(   Y^*_{k+1,1} \underline{v}_{0}^* \right)  
	~~C_{S} \right) }\vspace{-0.0cm}
$$ 
with  $C_{S}= \begin{pmatrix}
\underline{s}_{\ell_1}-\underline{s}^0_{g_1}&   \dots&   \underline{s}_{\ell_N}-\underline{s}^0_{g_N}
\end{pmatrix}^T$ and the   power constraint~\eqref{eqt:power_gen_car} 
can be rewritten  then as
$${ ~~~~~~~\forall~V \in \comp^N~~~~{ \begin{pmatrix}\star\\\star\\~\star~\end{pmatrix}}^* ~
	Q^{S_g}~
	{  	\begin{pmatrix}	V\otimes V^{*^T}\\V~~\\1~~	\end{pmatrix} }<0 }$$\\
with   
\begin{equation}
	Q^{S_g}=
	M_{S_g}^*\Psi  M_{S_g} -u_{N^2+N+1}^Tu_{N^2+N+1}
	\label{eqt: Q^{S_g}}
\end{equation}

\subsubsection*{Rewriting the current magnitude constraints}  
Using     current-voltage links \eqref{eqt:link_voltage_current}, the current   $\underline{i}_k$ is given by  \\ 
$$\underline{i}_{k}={   \left( 
	Y_{(k+1),1} ~~ \dots ~~ Y^*_{(k+1),N} ~\vline ~ Y _{(k+1),1}\underline{v}^*_0 
	\right) \begin{pmatrix} \underline{v}_1 \\  \vdots \\ \underline{v}_N \\\hline1  \end{pmatrix} }  $$
and the vector $\begin{pmatrix}\underline{i}_{1}&\dots&\underline{i}_{N}\end{pmatrix}^T$ rewrites then  as 
\begin{equation}
	\begin{pmatrix}\underline{i}_{1}&\dots&\underline{i}_{N}\end{pmatrix}^T= M_{I} \begin{pmatrix} \left( V\otimes V^{*^T }\right)^T~V^T~1 	\end{pmatrix}^T
	\label{eqt:new_forme_I} 
\end{equation}
with 
$$  M_I =    \begin{pmatrix} O_{N\times N^2}   & Y_{2:(N+1),2:(N+1)} & C_I \end{pmatrix} $$
where   $O_{N\times N^2}$ is the $N$ by $N^2$ null matrix,   $Y_{2:(N+1),2:(N+1)}$ is the sub-matrix of $Y$ which excludes the first row and the first column and  
$C_I=\begin{pmatrix} 	 Y_{2,1}\underline{v}_{0}&   \dots &   Y_{(N+1),1}\underline{v}_{0} 	\end{pmatrix}^T$.\\[3ex]
The current $\underline{i}_k$   can be  given by  
\begin{equation}\underline{i}_k=e_k~  \begin{pmatrix}\underline{i}_{1}&\dots&\underline{i}_{N}\end{pmatrix}^T
	\label{eqt:i_e_k_I}
\end{equation}
where $e_k\in \real^{N}$ is the $N$   null row vector     except the $k^\text{th}$  entry which is equal to 1.\\
The $N$ inequalities of~\eqref{eqt:current_limitation} rewrite as  
$$ \underline{i}_k^*~\underline{i}_k < \left( I_k^{max}\right)^2~~~~~~~~~k\in\mathcal{N}. $$
which can be rewritten, using  \eqref{eqt:new_forme_I}  and \eqref{eqt:i_e_k_I},  as\\
$${~~~~~\forall~ V  \in   \comp^N~~~{ \begin{pmatrix}\star\\\star\\~\star~\end{pmatrix}}^* 
	Q^{I}_k~ 
	{  	\begin{pmatrix}	V\otimes V^{*^T}\\V~~\\1~~	\end{pmatrix} } <0,~~k \in \mathcal{N}}$$
with  
\begin{equation}
	Q^{I}_k= M_I^*~e_k^*~e_k  M_I- \left( I_k^{max}\right)^2~u_{N^2+N+1}^Tu_{N^2+N+1} .
	\label{eqt:cuurent_QI}
\end{equation}



\subsection{Expressions of the different matrices   $\widetilde Q_\ell$   in Theorem~\ref{thm:feasp_with_polynomial_constraints}}
\label{Second_Appendix}
The objective of this appendix is to give the expressions of the different matrices   $\widetilde Q_\ell$   in Theorem~\ref{thm:feasp_with_polynomial_constraints}	which allow to characterize important  links between the different $X_k$ where $X_k$ is the $k^\text{th}$ element of     $X={\footnotesize\begin{pmatrix} \left( V\otimes V^{*^T }\right)^T~V^T~1 	\end{pmatrix}^T}$.  
Five different important links  appear
\begin{enumerate}[label=\subscript{L}{{\arabic*}}:]
	\item For every  four integers $a$, $b$, $c$ and $d$ taken in $\mathcal{N}$ 	
	$$ ~~~~~~~~\left(\underline{v}_a~ \underline{v}_b^* \right)^* \left(\underline{v}_c ~\underline{v}_d^* \right) = \left( \underline{v}_d~ \underline{v}_b^* \right)^* \left( \underline{v}_c ~\underline{v}_a^* \right) $$ 
	which means		
	$$ {  \left( X_{\scaleto{ (a-1)N+b}{7pt}}\right)^*  X_{\scaleto{(c-1)N+d}{7pt}}=  \left( X_{\scaleto{(d-1)N+b}{7pt}}\right)^*  X_{\scaleto{(c-1)N+a}{7pt}} }$$
	
		~\\

	\item For every  three integers $a$, $b$ and $c$ taken in $\mathcal{N}$  
	
	$$ ~~~~~~~~~~~~~\underline{v}_c^*~   \left(\underline{v}_b ~\underline{v}_a^* \right) = \underline{v}_a^*~   \left(\underline{v}_b ~\underline{v}_c^* \right)$$ 
	which means	
	
	$$~~~~~~~	 {  \left(  X_{\scaleto{N^2+c}{7pt}} \right)^* ~ X_{\scaleto{(b-1)N+a}{7pt}} = \left( X_{\scaleto{N^2+a}{7pt}}\right)^* ~X_{\scaleto{(b-1)N+c}{7pt}}}$$
	
	~\\
	
	\item For every  two  integers $a$ and $b$ taken in $\mathcal{N}$  $$ ~~~~~~~~~~~~~\underline{v}_c^*~   \left(\underline{v}_b ~\underline{v}_a^* \right) = \underline{v}_a^*~   \left(\underline{v}_b ~\underline{v}_c^* \right) $$
	which means		
	
	$$~~~~~~~~~~ 	 {  \left( X_{ \scaleto{(b-1)N+a} {7pt}  }\right)^* =  \left( X_{\scaleto{(a-1)N+b}{7pt}} \right)^*    }$$

		~\\

	\item For every    integer  $a$  in $\mathcal{N}$  	
	$$ ~~~~~~~~~~~~~2\left(\underline{v}_a~ \underline{v}_a^* \right)^*  =  \left( \underline{v}_a~ \underline{v}_a^* \right)+  \left( \underline{v}_a~ \underline{v}_a^* \right)^*  $$	 	 
	which means	
	
	$$~~~~~~~~~~ 	 {  2 \left( X_{ \scaleto{N^2+a}{7pt}}\right) =  X_{\scaleto{(a-1)N+a}{7pt}}  + \left( X_{\scaleto{(a-1)N+a}{7pt}}\right)^* }$$

		~\\

	\item For every    integer  $a$  in $\mathcal{N}$  	
	$$ ~~~~~~~~~~~~~\left(\underline{v}_a~ \underline{v}_a^* \right)  = \left(\underline{v}_a~ \underline{v}_a^* \right)^*$$
	which means		
	$$~~~~~~~~~~~~~~~~~ 	 {  X_{\scaleto{(a-1)N+a}{7pt}} = \left( X_{\scaleto{(a-1)N+a}{7pt}}\right) ^* }$$
	
\end{enumerate}

These equalities (in $X$) can be rewritten as   
$$ X^*~\widetilde{Q}_\ell~X=0 $$
where   $\widetilde{Q}_\ell$ is the $(N^2+N+1)$  by $(N^2+N+1)$   matrix full with zeros except few elements depending on the link.  

\begin{enumerate}[label=\subscript{L}{{\arabic*}}:]
	\item  For ${\tiny (a,b,c,d) \in    \mathcal{N} \hspace{-0.1cm}  \times \hspace{-0.05cm} \mathcal{N} \hspace{-0.1cm}  \times \hspace{-0.05cm}\mathcal{N} \hspace{-0.1cm}  \times \hspace{-0.05cm}\mathcal{N}}$, the elements of  $\widetilde{Q}_\ell$   are given by \\
	${ \left(\widetilde{Q}_\ell \right)_{i,j}=\left\lbrace \hspace{-0.2cm}
		\begin{matrix}
		&~~1 ~~~~\text{\footnotesize if $(i,j)=\left(b+N (a-1),d+N(c-1)\right) $}\\
		&~~1~~~~\text{\footnotesize if $(i,j)=\left( d+N(c-1),b+N(a-1)\right) $}\\
		&-1~~~~\text{\footnotesize if $(i,j)=\left( b+N(d-1),a+N(c-1)\right) $}\\
		&-1~~~~\text{\footnotesize if $(i,j)=\left(a+N(c-1),b+N(d-1)\right) $}\\
		&~~~~~~0~~~~\text{\footnotesize otherwise}~~~~~~~~~~~~~~~~~~~~~~~~~~~~~~~~~~~
		\end{matrix}\right.}$

	\item  For ${\tiny (a,b,c) \in    \mathcal{N} \hspace{-0.1cm}  \times \hspace{-0.05cm} \mathcal{N} \hspace{-0.1cm}  \times \hspace{-0.05cm}\mathcal{N} \hspace{-0.1cm}}$, the elements of   $\widetilde{Q}_\ell$  are given by \\
	${ \left(\widetilde{Q}_\ell \right)_{i,j}=\left\lbrace \hspace{-0.2cm}
		\begin{matrix}
		&~~1 ~~~~\text{\footnotesize if $(i,j)=\left(N^2+a,c+N(b-1)\right) $}\\
		&~~1~~~~\text{\footnotesize if $(i,j)=\left( c+N(b-1),N^2+a\right) $}\\
		&-1~~~~\text{\footnotesize if $(i,j)=\left( N^2+c,a+N(b-1)\right) $}\\
		&-1~~~~\text{\footnotesize if $(i,j)=\left(a+N(b-1),N^2+c\right) $}\\
		&~~~~~0~~~~\text{\footnotesize otherwise}~~~~~~~~~~~~~~~~~~~~~~~~~~~~ 
		\end{matrix}\right.}$
	
	\item  For ${\tiny (a,b) \in    \mathcal{N} \hspace{-0.1cm}  \times \hspace{-0.05cm} \mathcal{N} \hspace{-0.1cm}}$, the elements of   $\widetilde{Q}_\ell$   are given by   \\
	${ \left(\widetilde{Q}_\ell \right)_{i,j}=\left\lbrace \hspace{-0.2cm}
		\begin{matrix}
		&~~1 ~~~~\text{\footnotesize if $(i,j)=\left(N^2+N+1,b+N(a-1)\right) $}\\
		&~~1~~~~\text{\footnotesize if $(i,j)=\left(b+N(a-1),N^2+N+1\right) $}\\
		&-1~~~~\text{\footnotesize if $(i,j)=\left(a+N(b-1),N^2+N+1\right) $}\\
		&-1~~~~\text{\footnotesize if $(i,j)=\left(N^2+N+1,a+N(b-1)\right) $}\\
		&~~~~~~~0~~~~\text{\footnotesize otherwise}~~~~~~~~~~~~~~~~~~~~~~~~~~~~~~~~~~~ 
		\end{matrix}\right.}$
	
	\item  For ${\tiny a \in    \mathcal{N} \hspace{-0.1cm}}$, the elements of  $\widetilde{Q}_\ell$  are given by \vspace{-0.0cm} \\
	${ \left(\widetilde{Q}_\ell \right)_{i,j}=\left\lbrace \hspace{-0.2cm}
		\begin{matrix}
		&-1 ~~~~\text{\footnotesize if $(i,j)=\left(N^2+N+1,a+N(a-1)\right) $}\\
		&-1~~~~\text{\footnotesize if $(i,j)=\left(a+N(a-1),N^2+N+1\right) $}\\
		&~~~2~~~~\text{\footnotesize if $(i,j)=\left(N^2+a,N^2+a\right) $}~~~~~~~~~~~~~\\
		&~~~~~0~~~~\text{\footnotesize otherwise}~~~~~~~~~~~~~~~~~~~~~~~~~~~~~~~~~~ 
		\end{matrix}\right.}$

	\item  For ${\tiny a \in    \mathcal{N} \hspace{-0.1cm}}$, the elements of  $\widetilde{Q}_\ell$  are given by  \vspace{-0.0cm} \\
	${ \left(\widetilde{Q}_\ell \right)_{i,j}=\left\lbrace \hspace{-0.2cm}
		\begin{matrix}
		&~~\mathbf{j} ~~~~\text{\footnotesize if $(i,j)=\left(N^2+N+1,a+N(a-1)\right) $}\\
		&-\mathbf{j}~~~~\text{\footnotesize if $(i,j)=\left(a+N(a-1),N^2+N+1\right) $}\\
		&~~~~~~0~~~~\text{\footnotesize otherwise}~~~~~~~~~~~~~~~~~~~~~~~~~~~~~~~~~~~ 
		\end{matrix}\right.}$
	
\end{enumerate}

\medskip 

Please refer to Theorem~\ref{thm:feasp_with_polynomial_constraints} for   compact expressions  of the different matrices  $\widetilde{Q}_\ell$.

\subsection{Proof of  Theorem~\ref{thm:feasp_with_polynomial_constraints}}
\label{Third_Appendix}
	The pre and post multiplication of constraint~\eqref{eqt:main_eqt_thm_feasp_with_polynomial_constraints} by the vector  $X={  \left( \begin{smallmatrix} \left( V\otimes V^{*^T }\right)^T~V^T~1 	\end{smallmatrix}\right) ^T}$ results in 
$${ \hspace{-2cm}
	\begin{matrix}
	{ \begin{pmatrix}\star\\\star\\~\star~\end{pmatrix}}^* \hspace{-0.0cm} Q_0 ~
	{  	\begin{pmatrix}	V\otimes V^{*^T}\\V~~\\1~~	\end{pmatrix} }\hspace{-0.00cm} + \hspace{-0.0cm} \displaystyle \sum_{i=1}^{N+1} \hspace{-0.00cm} \tau_i { \begin{pmatrix}\star\\\star\\~\star~\end{pmatrix}}^* \hspace{-0.0cm} Q_i~
	{  	\begin{pmatrix}	V\otimes V^{*^T}\\V~~\\1~~	\end{pmatrix} }     \\[3ex]   ~~~~~~~~~~~~~~~~~~~~~~~~~~~ ~~~~~~~~~~~~~~~~ ~~~~~~~~~~~~~~~~~+ \displaystyle \sum_{j=1}^{N^E} \widetilde{\tau}_\ell~ { \begin{pmatrix}\star\\\star\\~\star~\end{pmatrix}}^*\hspace{-0.10cm}\widetilde{Q}_\ell~
	{  	\begin{pmatrix}	V\otimes V^{*^T}\\V~~\\1~~	\end{pmatrix} }  >0. 
	\end{matrix}}
$$
Given the form of the matrices  $\widetilde{Q}_\ell$, we obtain \\
$$~~~~ { \begin{pmatrix}\star\\\star\\~\star~\end{pmatrix}}^*\hspace{-0.0 cm}\widetilde{Q}_\ell~
{  	\begin{pmatrix}	V\otimes V^{*^T}\\V~~\\1~~	\end{pmatrix} }  =0 ~~~~~~~~~ \ell \in\{1,\dots,N^E\}$$\\
see Appendix~\ref{Second_Appendix} for more details. The last inequality rewrites then as \\[1ex]
$${ \hspace{+0.15cm}
	\begin{matrix}
	{ \begin{pmatrix}\star\\\star\\~\star~\end{pmatrix}}^* \hspace{-0.0cm} Q_0~ 
	{  	\begin{pmatrix}	V\otimes V^{*^T}\\V~~\\1~~	\end{pmatrix} }\hspace{-0.0cm} >- \hspace{-0.0cm} \displaystyle \sum_{i=1}^{N+1} \hspace{-0.00cm} \tau_i \hspace{-0.00cm} { \begin{pmatrix}\star\\\star\\~\star~\end{pmatrix}}^*\hspace{-0.0cm} Q_i~
	{  	\begin{pmatrix}	V\otimes V^{*^T}\\V~~\\1~~	\end{pmatrix} }     
	\end{matrix}}
$$\\[1ex]
Since     $\tau_i>0$ for $ i\in\{1,\dots,N+1\}$, the previous constraint results in  $$ { \begin{pmatrix}\star\\\star\\~\star~\end{pmatrix}}^* 
Q_0~
{  	\begin{pmatrix}	V\otimes V^{*^T}\\V~~\\1~~	\end{pmatrix} } >0 ~~~~~~~~~~~~~~~~~~~ $$ \\
for every  $V$ in $\comp^N$ satisfying
$$ 
~~~~~~~ { \begin{pmatrix}\star\\\star\\~\star~\end{pmatrix}}^* 
Q_i~
{  	\begin{pmatrix}	V\otimes V^{*^T}\\V~~\\1~~	\end{pmatrix} } <0, ~~  i\in\{1,\dots, N+1\} $$ 
which is the test to be performed  in Problem~\ref{prb:feasp_with_polynomial_constraints}. 

\subsection{Proof of  Corollary~\ref{cor:uncertain_power_flow_cor_1}}
\label{Fourth_Appendix}

	In Problem~\ref{prb:uncertain_power_flow_form_2},    after introducing the matrix ${\footnotesize \displaystyle \diag_{k=1,\dots,N}\left( {\left( V_k^{max}\right)^2}-{\left( V_k^{min}\right)^2} \right)}$ and since the  scalar $\vartheta$ is a positive constant,   the cost function in   Problem~\ref{prb:uncertain_power_flow_form_2} can be equivalently replaced by\vspace{-0.2cm}
	$$\min_{\left( V_k^{min}\right)^2,\left( V_k^{max}\right)^2}~~~ \textnormal{trace } {\displaystyle \left( \diag_{k=1,\dots,N}\left( {\left( V_k^{max}\right)^2}-{\left( V_k^{min}\right)^2} \right)\right) }$$
	subject to 
	$$~~~~~\begin{matrix}{ \begin{pmatrix}\star\\\star\\~\star~\end{pmatrix}}^* ~
	Q^{min}_k~
	{  	\begin{pmatrix}	V\otimes V^{*^T}\\V~~\\1~~	\end{pmatrix} } >0\\[4ex]
	{ \begin{pmatrix}\star\\\star\\~\star~\end{pmatrix}}^* ~
	Q^{max}_k~
	{  	\begin{pmatrix}	V\otimes V^{*^T}\\V~~\\1~~	\end{pmatrix} } >0 \end{matrix} ~~~~~  k\in\mathcal{N}$$ 
	for every  $V$ in $\comp^N$ satisfying \\
	$$ 
	~~~~~~~~~~~~~~~ { \begin{pmatrix}\star\\\star\\~\star~\end{pmatrix}}^* 
	Q_i~
	{  	\begin{pmatrix}	V\otimes V^{*^T}\\V~~\\1~~	\end{pmatrix} } <0, ~~  i\in\{1,\dots, N+1\} $$ \\
	Applying Theorem~\ref{thm:feasp_with_polynomial_constraints} with $Q_0=Q^{min}_k$ for  a fixed   $k$ in $\mathcal{N}$ results in \\
	$${ \hspace{1cm}
		\begin{matrix}
		&\displaystyle Q_k^{min}+ \sum_{i=1}^{N+1} \left( \tau^{min}\right)^k_i~Q_i +  \sum_{\ell=1}^{N^E} \left( \widetilde{\tau}^{min}\right)^k_\ell ~ \widetilde{Q}_\ell>0\\[3ex]
		&\left( \tau^{min}\right)^k_i>0~~~~~~~~~~~~~~ ~i\in\{1,\dots,N+1\}
		\end{matrix} } ~~~
	$$ 
	Thereafter,  rewriting these  constraints for every $k$ in $\mathcal{N}$,  using the functions  $\bdiag$ and $\diag$, results in   conditions~(i)  and~(iii) of  Corollary~\ref{cor:uncertain_power_flow_cor_1}. \\
	In the same manner, applying Theorem~\ref{thm:feasp_with_polynomial_constraints} with $Q_0=Q^{max}_k$ for  a fixed   $k$ in $\mathcal{N}$ results in \\
	$${ \hspace{1cm}
		\begin{matrix}
		&\displaystyle Q_k^{max}+ \sum_{i=1}^{N+1} \left( \tau^{max}\right)^k_i~Q_i +  \sum_{\ell=1}^{N^E} \left( \widetilde{\tau}^{min}\right)^k_\ell~  \widetilde{Q}_\ell>0\\[3ex]
		&\left( \tau^{max}\right)^k_i>0~~~~~~~~~~~~~~ ~i\in\{1,\dots,N+1\}
		\end{matrix} } ~~~
	$$ \\[1ex]
	and rewriting these  constraints for ever $k$ in $\mathcal{N}$    results in   condition~(ii)  and~(iv) of  Corollary~\ref{cor:uncertain_power_flow_cor_1}. \\
	Conditions~(v) and~(vi) are added to express  that the square of each bound  is positive and that    $ V^{max}_k > V^{min}_k$.\\[1ex]
	Please note that since Corollary~\ref{cor:uncertain_power_flow_cor_1} presents sufficient conditions,   only an upper bound   $\widetilde{ \mathscr{P}_{opt}}$ on the   optimal perimeter ${ \mathscr{P}_{opt}}$ of Problem~\ref{prb:uncertain_power_flow_form_2} can be obtained.   
 


\addtolength{\baselineskip}{-2pt}
\bibliographystyle{ieeetr}
\bibliography{ifacconf}

\begin{thebibliography}{10}

\bibitem{FVHA:13}
M.~Fan, V.~Vittals, G.~T. .Heydt, and R.~Ayyanar, ``Probabilistic power flow
  analysis with generation dispatch including photovoltaic resources,'' {\em
  IEEE Trans. Pow. Systems}, vol.~28, no.~2, pp.~1797--1805, 2013.

\bibitem{BCH:14}
D.~Bienstock, M.~Chertkov, and S.~Harnett, ``Chance-constrained optimal power
  flow: Risk-aware network control under uncertainty,'' {\em SIAM Review},
  vol.~56, no.~3, pp.~461--495, 2014.

\bibitem{ChD:17}
M.~Chertkov and Y.~Dvorkin, ``Chance constrained optimal power flow with
  primary frequency response,'' in {\em IEEE Conf. Decision and Control},
  (Melbourne), pp.~4484--4489, Dec. 2017.

\bibitem{LZSWYG:14}
L.~Luo, J.~Zhu, S.~Yang, K.~Wang, J.~Yao, and W.~Gu, ``Interval arithmetic
  based influence analysis on power flow caused by integration of wind power
  and electric vehicles,'' {\em IFAC Proceedings Volumes}, vol.~47, no.~3,
  pp.~2758 -- 2763, 2014.

\bibitem{LLZWQ:17}
X.~Liao, K.~Liu, Y.~Zhang, K.~Wang, and L.~Qin, ``Interval method for uncertain
  power flow analysis based on {T}aylor inclusion function,'' {\em IET
  Generation, Transmission Distribution}, vol.~11, no.~5, pp.~1270--1278, 2017.

\bibitem{SaS:06}
S.~Saric and A.~Stankovic, ``An application of interval analysis and
  optimization to electric energy markets,,'' {\em IEEE Trans. Pow. Systems},
  vol.~21, pp.~515--523, May 2006.

\bibitem{CJD:11}
Y.~C. Chen, X.~Jiang, and A.~D. Dom\'{i}nguez-Garc\'{i}a, ``Impact of power
  generation uncertainty on power system static performance,'' in {\em 2011
  North American Power Symposium (NAPS)}, pp.~1--5, Aug 2011.

\bibitem{BEFB:94}
S.~Boyd, L.~{El~Ghaoui}, E.~Feron, and V.~Balakrishnan, {\em Linear Matrix
  Inequalities in Systems and Control Theory}, vol.~15 of {\em Studies in
  Appllied Mathematics}.
\newblock Philadelphia, {USA}: {SIAM}, Jun 1994.

\bibitem{Ulf:01}
U.~J\"onsson, {\em Lecture notes on integral quadratic constraints}.
\newblock Royal Institute of Technology {(KTH)} Stockholm, Sweeden, 2001.

\bibitem{LKMS:18}
K.~Laib, A.~Korniienko, F.~Morel, and G.~Scorletti, ``{LMI} based approach for
  power flow analysis with uncertain power injection,'' {\em The joint 9th
  {IFAC} {S}ymposium on {R}obust {C}ontrol {D}esign (ROCOND) and 2nd {IFAC}
  {W}orkshop on {L}inear {P}arameter {V}arying {S}ystems (LPVS)}, Sept. 2018.
\newblock (to appear).

\bibitem{Par:03a}
A.~P. Parrilo, ``Semidefinite programming relaxations for semialgebraic
  problems,'' {\em Mathematical Programming}, vol.~96, no.~2, pp.~293--320,
  2003.

\bibitem{MaP:15}
A.~R. Malekpour and A.~Pahw, ``Radial test feeder including primary and
  secondary distribution network,'' in {\em 2015 North American Power Symposium
  (NAPS)}, pp.~1--9, Oct 2015.

\end{thebibliography}
\addtolength{\baselineskip}{2pt}

\end{document}